# RATIONALE AWARENESS FOR QUALITY ASSURANCE IN ITERATIVE HUMAN COMPUTATION PROCESSES

Lu Xiao

The University of Western Ontario
North Campus Building 256
London, Ontario, Canada
e-mail: lxiao24@uwo.ca

## ABSTRACT

Human computation refers to the outsourcing of computation tasks to human workers. It offers a new direction for solving a variety of problems and calls for innovative ways of managing human computation processes. The majority of human computation tasks take a parallel approach, whereas the potential of an iterative approach, i.e., having workers iteratively build on each other's work, has not been sufficiently explored. This study investigates whether and how human workers' awareness of previous workers' rationales affects the performance of the iterative approach in a brainstorming task and a rating task. Rather than viewing this work as a conclusive piece, the author believes that this research endeavor is just the beginning of a new research focus that examines and supports meta-cognitive processes in crowdsourcing activities.

## INTRODUCTION

Human computation refers to the outsourcing of computation tasks to human workers. For example, in the ESP Game, human computation power is leveraged to address the computation problem of labeling images with metadata. Wikipedia, the largest online encyclopedia, is a classic product of human computation processes. Amazon's Mechanical Turk provides a platform for human workers to get paid for completing tasks. In human computation processes, tasks are distributed to human workers who work on the task either in parallel (e.g., Threadless) or through an iterative process (e.g., collaborative writing in Wikipedia). To understand the tradeoffs between these two approaches, Little et al. (2010) ran a series of controlled comparative experiments for three different types of tasks using Amazon's Mechanical Turk, and examined the effectiveness of these two approaches for each task. Their experiments suggested that the iterative approach has potential benefits for brainstorming tasks in human computation processes. Little et al. (2010) called for more research work to investigate the use of the iterative approach and how to coordinate workers' work in human computation processes.

In response to Little et al.'s call (2010), this paper reports a study that explored whether and how workers' awareness of previous workers' rationales affected the quality of ideas in brainstorming tasks. A rationale is an explanation of the reasons underlying decisions, conclusions, and interpretations. Prior studies have shown that awareness of other members' rationales supports one's associative thinking process in group work, helping one generate more ideas (Xiao, 2011a). It is therefore hypothesized that in a brainstorming task that is completed through human computation processes, showing previous workers' rationales of their ideas (i.e., why they generated the ideas) improves the quality of the iterative approach by supporting the workers' associative thinking process.

The study also examined whether presenting the rationales behind the generated ideas affected the evaluation of these ideas in human computation processes. A rationale explains why the worker believes the idea generated is appropriate for the brainstorming task. Presenting this rationale to all the raters of the idea provides the raters with the same information related to the evaluation task. It is thus hypothesized that doing so will reduce the variation between evaluations by multiple raters.

A comparative experiment was conducted to test these hypotheses. For a better comparison between the results of this experiment and Little et al.'s work (2010), the study adopted Little et al.'s model on iterative processes and adapted their experimental design for brainstorming tasks (2010).

## RELATED WORK

### Human Computation Processes

Few research studies have been conducted to understand coordination of work in human computation processes. Quinn and Bederson (2009) identified a taxonomy to describe distributed human computation processes. The aggregation dimension of the taxonomy addresses how work is coordinated and combined to achieve a final result. Malone et al. (2009) differentiated the types of tasks and working modes of human workers. They identified three conditions for collective intelligence: Collection, Contest, and Collaboration. Collection condition refers to the situation where the task can be divided into small components that can be done independently of each other most of the time. Contest condition is when only one or a few of the solutions generated by the crowd will be needed. Collaboration condition is when the activity requires interdependent work and cannot be decomposed into independent pieces.

The goal of Little et al.'s study (2010) was to explore the benefits of showing the current worker the content previously generated by other workers. Little et al. (2010) presented two models to depict human computation processes: iterative and parallel. The iterative model consists of a sequence of creation tasks such that the result of each task becomes an input to the next one. In the parallel model, although the workers are doing the same kind of task as in the iterative model, no workers are shown any work created by others.

Comparing Malone et al.'s conditions (2009) with Little et al.'s model of human computation processes (2010), the Collection condition is similar to the parallel approach, whereas the Collaboration condition is closer to the iterative approach.

To explore the iterative and parallel approaches in human computation processes, Little et al. (2010) conducted a series of experiments investigating tradeoffs between each approach in three crowdsourcing tasks: writing, brainstorming, and transcription. They found that the iterative approach has statistically significant better performance than the parallel approach in writing and brainstorming in terms of the average quality of responses. However, in brainstorming and transcription tasks, the responses in the iterative approach tend to converge, with fewer best quality responses. Although the parallel approach has more occurrences of best quality responses, the average quality is lower than in the iterative approach because of the high variation of the responses. That is, the iterative approach produced a better quality with respect to the average rating of the results, whereas the majority of the responses with best rating appeared in the parallel approach. Little et al. (2010) called for future investigations on ways to help generate the best ideas in the iterative approach.

### Rationale Sharing in Virtual Group Activities

A rationale is an explanation of the reasons underlying decisions, conclusions, and interpretations. Prior studies have shown that in group settings, sharing rationales is important to the success of group activities and critical in group processes (e.g., Conklin and Yakemovic, 1991; MacLean et al., 1991). In her study of how articulating and sharing rationales within the group affected group brainstorming activities in a virtual environment, Xiao found that one's awareness of other members' rationales (i.e., rationale awareness) supported one's awareness of his/her domain knowledge and intellectual contribution (Xiao, 2011b, 2011c). Xiao's study also indicated that one's rationale awareness supported his/her associative thinking process (Xiao, 2011a) – an important process for generating new ideas – and helped to monitor and control the quality of the group work (Xiao, 2011c).

## RESEARCH DESIGN

In response to Little et al.'s (2010) call to explore ways of improving the performance of the iterative approach in human computation processes, this study investigates the usefulness of showing the rationales of previous workers' decisions in generating good quality ideas in brainstorming tasks. Based on the existing knowledge about the effects of sharing rationales and rationale awareness in group activities, the first hypothesis of the study is:

*Hypothesis 1: In a brainstorming task, the quality of the iterative approach will be improved by making the current worker aware of previous workers' rationales for their ideas (i.e., why they suggested the ideas).*

A rationale explains why the worker believes his/her idea is a good one for the given brainstorming task. This provides workers who evaluate the idea with more contextual information about the idea than the brainstorming task context alone. Making this rationale available to all evaluators of the idea implicitly offers a shared understanding among them about the idea, potentially reducing the variations between their evaluations. The second hypothesis is therefore:

*Hypothesis 2: Variations between multiple evaluations will be reduced by making the rationale behind the idea available to all workers.*

An experiment was conducted to test these hypotheses. The brainstorming task and the rating task from Little et al.'s (2010) experiment were used, as well as elements of their experimental design, such as the number of brainstorming and rating tasks and the ordering of the tasks. Also, similar to Little et al.'s experiment (2010), Amazon's Mechanical Turk was used as the platform to support these tasks. Turkit was used to develop iterative human computation tasks. Workers in Amazon's Mechanical Turk are called turkers. Hence the term "turkers" is used in the remaining sections to refer to people who participated in the experiments.

The details of the tasks and the results are presented below.

### BRAINSTORMING TASK

Similar to the brainstorming task in Little et al.'s (2010) work, the brainstorming task in this experiment required each turker to generate five company names for the provided company description. But as our experiment examined the iterative approach exclusively, the names generated by previous turkers were available to the current turker in all brainstorming tasks in the experiment. One iteration of a brainstorming task required one turker to generate five company names for the given company description, and each iteration was counted as one HIT (Human Intelligence Task) in Mechanical Turk. Another difference is that the brainstorming task of this study required turkers not only to generate the company names but also to provide a rationale for each name to explain why it was a good name for the company description.

Depending on the condition of the brainstorming task, previous turkers' rationales were or were not shown to the current turker. The task interface for the condition of showing previous turkers' names and rationales is presented in Figure 1.

**Brainstorming company names given company description**

**Note:** If you have worked on this company description task already, please don't work on it in this task; otherwise, you won't be paid for doing the same company description task twice. if you have generated five names for this company description for us before, you cannot participate in this HIT any more. Or you won't be paid for this work. You may see some names created by other people, please **don't enter the same name given by other people**; otherwise you won't be paid. Besides the names, you should also provide a rationale. A rationale that simply states the name fits the description will not be counted. Examples of a bad rationale would be "It fits your description", "it shows your description", "it matches with the description". Don't use this kind of rationale, otherwise, you will not be paid.

**Company Description:**
A good watch is a symphony of precision. Time isn't stopping so it's sensible to keep track of it. Wrist watches can be stylish, light-weight and always on display for others to see. Are you planning to go to the seaside? We have a waterproof watch for that. Come to our watch store where luxury costs less.

| Provide 5 different company names below | Briefly explain why this is a good company name? (max 40 characters) |
|---|---|
|  |  |
|  |  |
|  |  |
|  |  |
|  |  |

submit

| Company names given by other people | Corresponding rationales |
|---|---|
| Watching Time | A play on word :) |
| Time Marches on | The sons makes me think of watches |
| Timely Gifts | goes with the watch idea |
| B on time | why people buy a watch |

*Figure 1: Turkers were asked to generate five new company names given the company description. In the control condition, the interface showed only the names suggested so far. In the treatment condition, the interface showed the rationales of the suggested names as well*

As in Little et al.'s (2010) experiment, six different fabricated company descriptions were used in the experiment, and each company description had six iterations. In the experiment, three company descriptions had one condition posted first then the other next, and the other three company descriptions had these conditions posted in reverse order.

Because quality control of the results was important for the experiment, several criteria were added:

- A HIT result should have rationales for all five generated names
- A HIT result should not have any known brand name
- A HIT result should not have the same names generated by previous turkers (note: they were made available to the current turker in this iterative approach)
- To participate in the experiment, a turker should have a Mechanical Turk approval rate of at least 97%.

To reduce the noise and bias introduced by turkers, the task description stated that a turker could participate in the brainstorming task only once for the particular company description. However, as in Little et al.'s (2010) experiment, a turker was allowed to participate in different brainstorming tasks for different company descriptions.

After all iterations were completed for a company description, the turkers' results were examined. Although all the criteria were stated in the task description (see Figure 1), there were still disqualified results that had to be discarded in the rating task. When an iteration was removed, all its generated names and rationales were removed from the experiment, and hence that iteration was lost. To make up for it, a brainstorming task was posted again so as to add one iteration. Names and rationales from the discarded iteration were not presented to the newly added iteration. Table 1 presents the iterations discarded in each brainstorming task. In total, 180 unique names were generated in the condition of showing previously generated names only. In the condition of showing previously generated names and their rationales, the turker who worked for the second company description for the 3rd iteration generated four qualified names and rationales but had the rationale missing for the last name. His/her results were kept in the analysis, except for the last name. Therefore, there were 179 unique names generated in this condition.

| Brainstorming Condition | Company Descriptions | | | | | |
|---|---|---|---|---|---|---|
| | 1 | 2 | 3 | 4 | 5 | 6 |
| Previously generated names were shown to the current turker | $1^{st}$, $2^{nd}$ | $4^{th}$, $5^{th}$, and $6^{th}$ | $3^{rd}$ | $4^{th}$ | | $3^{rd}$ |
| Previously generated names AND rationales were shown to the current turker | $5^{th}$, $6^{th}$ | $1^{st}$ | $3^{rd}$, $6^{th}$ | $2^{nd}$ | $1^{st}$, $4^{th}$, and $5^{th}$ | $2^{nd}$, $3^{rd}$ |

*Table 1: Original iterations that were discarded from the experiment*

**Rating Task**

After the brainstorming task was completed for all six company descriptions, a rating task was posted on Mechanical Turk to evaluate the quality of the generated company names. In this task, each turker was asked to score a company name, given the company description. There were also two conditions in the rating task: showing vs. not showing the name's rationale to the rater. Figure 2 presents a screenshot of a rating task that had the name and the rationale shown to the rater. Each rating task required turkers to rate one name for a company description in one condition. The rating task was posted as a HIT in Mechanical Turk. It had 10 assignments for turkers. In other words, each name in a rating condition was rated by 10 different turkers.

Turkers were not allowed to rate the same name in both rating conditions, or names they had created in brainstorming tasks. But they were allowed to rate different names in two conditions, and they could rate multiple different names for the same company description in both conditions.

Table 2 shows the design and the amount of data for analysis in the experiment.

**Rating Company Name**

Note:
- Rate the company name below according to the company description. Score 10 is the best and score 1 is the worst.
- If you have worked on this company description in the brainstorming task already, please don't work on it in this task; otherwise, you won't be paid.
- We will check the ratings, if any irresponsible rating is found, it will cause the rejection for all the HITs completed by the same person. Please be very responsible!!

Company Description:
A good watch is a symphony of precision. Time isn't stopping so it's sensible to keep track of it. Wrist watches can be stylish, light-weight and always on display for others to see. Are you planning to go to the seaside? We have a waterproof watch for that. Come to our watch store where luxury costs less.

Score 10 is the best and score 1 is the worst

| Name | Rationale | Score |
|---|---|---|
| watch life | watch life in your watch | ○1 ○2 ○3 ○4 ○5 ○6 ○7 ○8 ○9 ○10 |
| | submit | |

*Figure 2: Turkers rated a company name ("watch life" in this screenshot) given the company description. In the control condition, the interface showed only the name. In the treatment condition, the interface showed the rationale of the suggested name as well*

| | | Rating Task | |
|---|---|---|---|
| | | **Control condition** (rationale of the name not shown) | **Treatment condition** (rationale of the name shown) |
| **Brainstorming Task** | control condition (rationales of previously generated names not shown) | 1800 ratings for 180 unique names | 1800 ratings for 180 unique names |
| | treatment condition (rationales of previously generated names shown) | 1790 ratings for 179 unique names | 1790 ratings for 179 unique names |

*Table 2: The design and data of the first experiment*

### RESULTS

In the analysis, a name's rating in a rating condition is the average of its 10 ratings in that condition, and has a standard deviation that tells the variation of the 10 ratings. Minitab 16 was used for statistical analysis.

### The Effect of Showing the Name's Rationale in the Name's Evaluation

In the rating task, each rater was required to judge the quality of the name. So conceptually speaking, a rating should not be affected by whether a rationale was shown or not. However, the rhetorical effect of a rationale may "convince" the rater that the name had a better/worse quality than the rater thought, and hence it may have an impact on the rating. The result of a paired t-test for the ratings from the two rating conditions yielded a p-value of .57 (t = -.57), which indicated that there was no statistically significant difference between the two rating conditions on the names' ratings.

### Hypothesis 1

Little et al.'s (2010) approach of examining the average and best quality of the generated names was used in testing the first hypothesis. Specifically, the average rating of all iterations in a brainstorming condition was calculated by averaging ratings of all the names generated from that iteration for all six companies. For better comparison between Little et al's results and ours, a name's ratings only included those from the control condition of the rating task. Figure 3 plots these average ratings of the two brainstorming conditions. Note that Little et al.(2010)

presented a similar figure to compare the average quality of generated names in parallel and iterative conditions.

Surprisingly showing the rationales in the brainstorming task did not improve the average rating of names all for the iterations.

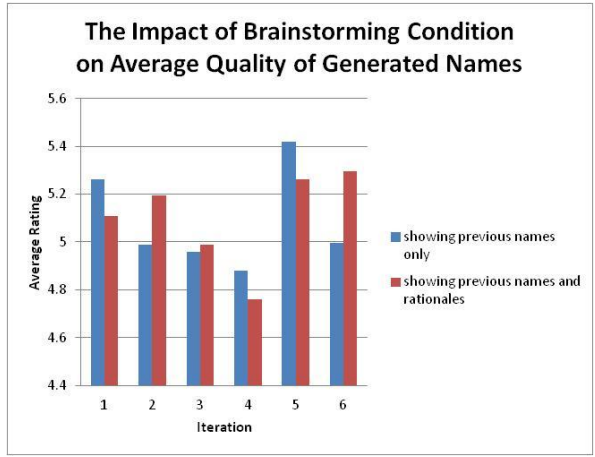

*Figure 3: The iterations' average rating in two brainstorming tasks*

In Little et al.'s experiment (2010), the authors claimed that although the iterative approach improved the average quality of generated names, the parallel process generated the best rated names. The data they presented to support the claim included: 1) 4 of the 6 fake companies had best rated names in the parallel approach; and 2) one company had a best rating of 7.3 in the iterative approach vs. 8.3 in the parallel approach.

To enable a comparative analysis between the two studies, the best rated names in both brainstorming conditions in the current study were examined. It was found that the condition of showing rationales in the name generation task did not improve the best quality of the generated ideas (Table 3).

| Company No. | Brainstorming Condition | | Difference |
| --- | --- | --- | --- |
| | Showing previous names only | Showing previous names and rationales | |
| 1 | 7.2 | 7.2 | 0 |
| 2 | 7 | 6.5 | -.5 |
| 3 | 6.7 | 6.4 | -.3 |
| 4 | 6.7 | 6.8 | .1 |
| 5 | 7 | 6.4 | -.6 |
| 6 | 6.2 | 7 | .8 |

*Table 3: Best ratings of six fake company descriptions in two brainstorming conditions*

Given that the difference between ratings of best rated names in the two conditions is quite small, and given the limited data provided in Little et al.'s paper (2010), it is not wise to conclude that showing previous rationales in the iterative approach improves the best quality of generated names in the iterative approach. More investigation is needed to examine this hypothesis.

### Hypothesis 2

A paired t-test for the standard deviations of the ratings of the two rating conditions was conducted. It was found that showing rationales reduced variations statistically significantly ($t_{359}= -3.49$, $p = 0.001$).

However, the standard deviation was larger in the rating condition of showing rationales (2.53 vs. 2.43), which contradicted the hypothesis. This seemed to suggest that although additional relevant information was shared among raters, the lack of grounding process made the assumption of shared understanding invalid. Because raters had different understandings regarding the information, this actually created another dimension for variation of evaluation, thus increasing the variation of the evaluation result as opposed to decreasing it.

### DISCUSSION

We identified three issues in our experimental design that could have impacted the results. Similar to Little et al.'s experimental design (2010), our experiment allowed a turker to participate in a brainstorming task for different company descriptions. However, this feature may have impacted our experimental results more significantly than those of Little et al (2010). When turkers worked on a different description in our experiment, the fact that they had articulated their rationales in previous brainstorming tasks may have had a greater influence on the current task than was the case in Little et al.'s experiment. This is because when articulating one's rationales, one may be engaged in a reflective thinking process, and over time may become better at reflecting on one's choices (e.g., the generated names) and may even change one's choices after reflection (Xiao et al., 2008; Xiao, 2011a, 2011b). So if a turker has generated five names for a company description, he/she has gone through the "training" of reflective thinking five times. This then could have affected his/her way of thinking when engaged in the name generation task for a different company description. Table 4 presents the number of unique turkers in each task for different experiment conditions. As shown in the table, a fairly high proportion of turkers in the brainstorming task (16.67% for rationales not shown; 30.56% for rationales shown) worked on multiple tasks for different companies. This effect may be subtle, and it may not be statistically significant. However, in future research into human computation processes it will be important to check the number of unique turkers, especially in experiments in which the design interventions may have similar learning effects that can carry over from task to task.

|  | Experiment Condition | The number of unique turkers in the experiment | The possible number of unique turkers |
|---|---|---|---|
| Brainstorming Task | Rationales were *not* shown to the current turker | 30 | 36 |
|  | Rationales were shown to the current turker | 25 | 36 |
| Rating Task | The name's rationale was *not* shown to the current rater | 69 | 3590 |
|  | The name's rationale was shown to the current rater | 99 | 3590 |

*Table 4: The number of unique turkers for different tasks*

The second issue refers to the procedure of running the six iterations and discarding the disqualified HITs. In the experiment, six iterations were completed before the generated HITs were examined for approval/disapproval. In addition, new iterations were added to make up for the discarded iterations (see Table 1). Although the names and rationales generated from discarded iterations were not presented in the later iterations, the accepted iterations were influenced by the discarded iterations, as turkers of the accepted iterations had already seen the names and rationales that were removed. Thus their performance was to some extent affected by the awareness of these discarded names and rationales. For example, a turker who had seen previous names such as Rolex or Timex for the watch company description may have been discouraged from generating good names and providing good rationales, since previous turkers were not being serious in the task. This influence should be taken into account. It is recommended that in future experiments investigating the iterative approach, the quality of the HITs for the current iteration should be judged before moving on to the next iteration.

The third issue refers to the number of company descriptions used in the experiment. As in Little et al. (2010), the effects of company descriptions were not examined in the analysis, except when best rated names were analyzed (Table 3). The assumption was that, in terms of the effects of individual differences and replicability of the experiment, each company description was analogous to an individual participant in a controlled lab study. Given the results, it appears that six company descriptions were not enough. And to compare the results between the brainstorming conditions, six iterations were used, and each iteration had five names generated by a turker. We used six company descriptions so as to compare the findings with Little et al.'s work (2010), which also used these parameters in their experiment. But the question is, are these parameters good enough for comparison? It is possible that some observations made here could not be generalized, as the number of company descriptions is too small if we consider it to be analogous to the number of participants in a traditional experiment. It is also possible that the number of iterations was not big enough to confirm the differences, e.g., the slightly better rated names in the rationale sharing condition. Perhaps one next step is to examine these parameters in a more rigorous way to understand what factors matter among different approaches in human computation processes with respect to management, coordination, and quality control.

**CONCLUSION**

This paper reports a study about the effects of showing previous turkers' rationales in human intelligence tasks posted on Amazon's Mechanical Turk. The research design adapted Little et al.'s (2010) approach, but focused on brainstorming tasks only. The results we obtained from this experiment were not expected. To better interpret the results, we reflected on several issues related to the experiment design and quality control of iterative human computation processes. It is believed that by examining the quality of HITs after all the iterations were completed had important impact on the results: bad rationales from previous iterations may have had negative impact on the subsequent quality of generated ideas.

At the time of this submission, we were in analysis of the second experiment that used the same experiment design and the same company descriptions. There were two differences in the second experiment: parallel condition was included in the brainstorming task (it also required turkers to provide the rationales of the ideas); and rationale quality was examined at the end of each iteration before moving to the next iteration. The results of the second experiment were different from this experiment: showing the name's rationales improved the average quality of the generated ideas; and providing idea's rationales in the rating tasks reduced the variation of the idea's ratings. The comparison of the results from these two experiment suggested that previous HITs' rationale quality had big impact on the subsequent HITs, providing an evidence of our interpretation of the first experiment's results. One consistent finding in the second experiment is that showing previous HITs' rationales did not improve best quality of generated ideas in the iterative human computation processes. The details of the second experiment can be found at (Xiao, 2012).

These findings are contextualized and may be bounded by parameters chosen in the experiment. More investigation is needed to examine the effects of showing other workers' rationales in human intelligence tasks in the iterative approach. For example, as a reviewer of the paper suggested, the accumulative effect of showing previous turkers' rationales was not considered in this experiment. What if the system allowed the current turker to pass on the good names and good rationales to the next turker and provide his/her rationale of why these names and rationales are good? How will such accumulative effect impact the quality of the work? Will such system allow the current turker not only gain financial benefit, but also practice evaluation and reasoning skills?

Rather than viewing this work as a conclusive piece, the author believes that this research endeavor is just the beginning of a new research focus that examines and supports meta-cognitive processes in crowdsourcing activities.

## ACKNOWLEDGEMENT

I thank Da Kuang for programming the tasks using Turkit, and the turkers who chose to do the tasks. I thank MIT researchers for making Turkit open source software.